\documentclass[twocolumn,showplaces,preprintnumbers,amsmath,amssymb,prl,aps]{revtex4-1}

\usepackage{epsfig}
\usepackage{graphicx}
\usepackage{bm}
\usepackage{float} 
\usepackage{amsmath}
\usepackage{gensymb}
\usepackage{epstopdf}

\begin{document} 

\title{Evidence of local excitons in the core level spectra of Si/Ge inverted quantum hut embedded silicon}


\author{Arka Bikash Dey$^{1}$, Milan K Sanyal$^1$, Swapnil Patil$^2$, Khadiza Ali$^2$, Deepnarayan Biswas$^2$, Sangeeta Thakur$^2$, and Kalobaran Maiti$^2$}



\altaffiliation{Corresponding author: kbmaiti@tifr.res.in}
\affiliation{
 $^1$Surface Physics and Material Science Division, Saha Institute of Nuclear Physics, Kolkata 700064, INDIA \\
 $^2$Department of Condensed Matter Physics and Materials Science, Tata Institute of Fundamental Research, Homi Bhabha Road, Colaba, Mumbai - 400005, INDIA}


\begin{abstract}
Conversion of Si to a direct bandgap semiconductor for optoelectronic application is a great challenge for many decades. It is proposed that embedment of suitable sized quantum dots into silicon matrix may be exploited to convert silicon to a direct bandgap semiconductor. The other bottleneck to this outstanding issue is the identification of local excitons, a signature of direct bandgap property and their comportment within the dots that can be utilized in engineering optoelectronic devices, quantum communications, etc. We studied the core level spectra of Si/Ge quantum huts embedded Si employing high resolution photoemission spectroscopy. Inverted quantum huts (IQHs) of Ge (13.3nm $\times$ 6.6nm) were grown on a Si buffer layer deposited on Si(001) surface using molecular beam epitaxy method and the photoemission experiments were carried out at different locations of the IQH structures exposed via controlled sputtering and annealing processes. We discover distinct features in the Ge 3$d$ core level spectra at the lower binging energy side of the bulk 3$d$ peak in contrast to the scenario of core level satellites often observed due to photoemission final state effects. The energy of these features are found to be sensitive to the location of the IQH structure probed revealing different core hole screening by the excitons located at different parts of IQHs. These results reveal local character of the excitons in the IQHs necessary for type I photoluminescence and establish core level spectroscopy as a direct probe of local excitons. These finding are expected to help amalgamation of microelectronics and solid state photonics important for optoelectronic applications.
\end{abstract}


\maketitle

Advances in silicon integrated circuits (ICs) based technology \cite{Riordan99,Meindl02} has led to well-developed microelectronic devices containing more than two billion transistors on a single chip that satisfies Moore’s law\cite{Schaller97} over several decades. Silicon, an indirect bandgap semiconductor, is a poor optical material requiring phonon assistance for electronic transitions \cite{Saito98}. Nanostructure embedment in Si \cite{Priolo14} is being exploited to engineer high-speed low-power optical output devices\cite{Cho04},
light emitters\cite{Chan01}, lasers\cite{Rong05}, electro-optic modulators\cite{Xu05}, etc., where the photoluminescence yield is comparable to the direct bandgap quantum dots\cite{Cao10}. Recent experiments show carrier multiplication efficiency of 190\% in germanium nanocrystal of 5-6 nm size \cite{Saeed15} and quantum-confined exciton induced strong Stark effect in thin germanium quantum-well on silicon \cite{Kuo05}. However, the strain at the Si/Ge interface generally gives rise to type-II band alignment \cite{Kurdi06} exhibiting power law dependence (exponent = 0.7 - 1.5) \cite{Ueda12} of photoluminescence energy making it an inefficient optical material. A density functional theoretical study shows signature of direct band gap in Ge quantum structure grown on Si(001) and Si(111) crystals, while the indirect bandgap survives for Si(110) substrate \cite{Kholod00}. Moreover, incorporation of Ge into Si matrix enhances hole mobility \cite{Wang89} and photoluminescence emission near 1.5 $\mu$m region\cite{Suess13}. Thus, SiGe quantum structures embedded in Si appears to be a promising candidate for optoelectronic applications.

\begin{figure}
\includegraphics[scale=0.4]{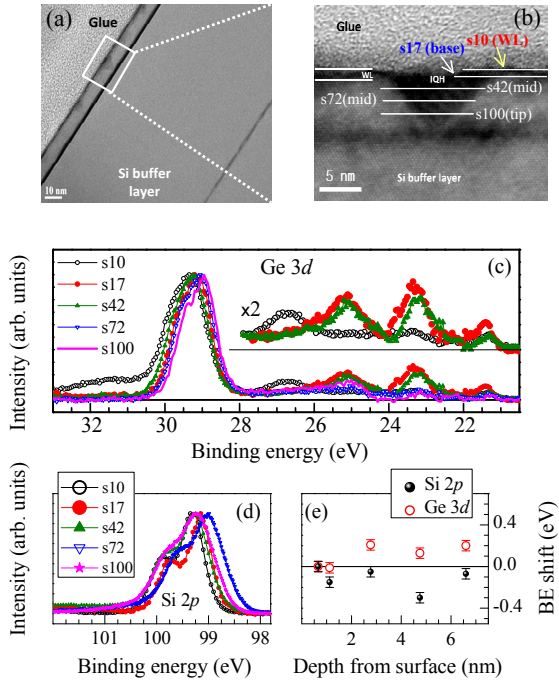}
\vspace{-2ex}
\caption{(a) XTEM image of Si/Ge quantum structure; dark and grey colours represent Ge and Si atoms, respectively. (b) High resolution XTEM image of the boxed region in (a). The horizontal lines represent the exposed surface after sputter anneal cycle; `s10(WL)' refers to 10 minutes sputtered surface revealing the wetting layer (WL). At s17, s42, s72 and s100, the exposed surfaces are the base region (depth 1.1 nm), mid region (depth 2.7 nm), mid region (depth 4.7 nm) and tip of IQHs (depth 6.6 nm), respectively. (c) Ge 3$d$ core level spectra exhibiting new features in energy range 21-28 eV. The inset shows 21-28 eV region with rescaled intensity. (d) Si 2$p$ core level spectra exhibiting energy shift at different exposed surfaces. (e) The energy shift of the Ge 3$d_{5/2}$ (open symbols) and Si 2$p_{3/2}$ (solid symbols) peaks at different locations of the quantum huts relative to s10 peaks.}
\label{Fig1-core}
\end{figure}

In order to address the outstanding issue of the behaviour of excitons \cite{Koch06}, we employed core level photoemission spectroscopy (CoLePES), where incident photons excite core electrons at a fast time scale ($\sim$ attosecond range) revealing the element specific local electronic structure. Small escape depth of the photoelectrons makes the technique surface sensitive\cite{surf}. We have exploited these features to build a platform for studying the behaviour of excitons located at different parts of the Si/Ge quantum structures.

Ge quantum structures were grown on a Si buffer layer prepared on Si(001) surface using molecular beam epitaxy, where the Ge atoms diffuse into the Si buffer layer. Quantum dots are usually prepared employing Stranski-Krastanov (S-K) mode-based epitaxial growth of self-assembled nanostructures, where the three dimensional islands are formed at the surface/interface followed by two dimensional wetting layer (WL) growth for II-V, III-IV and IV-IV semiconductor materials. Instead, here, Ge atoms diffuse into the underneath Si buffer layer grown on the substrate at temperatures of about 430 $^{\circ}$C and form special type of self-protected \emph{inverted quantum huts} (IQHs) \cite{growth,Sharma14,Sharma15}; IQHs of much bigger size (75nm base and 21nm height) are reported to form at lower growth temperatures \cite{Soo01}. The high resolution XTEM (cross sectional tunneling electron microscopy) image shown in Fig. \ref{Fig1-core} exhibits a dimension of 13.3 nm (base) $\times$ 6.6 nm (depth) with sharp boundary; the confined volume is significantly smaller than the previously reported SiGe IQHs\cite{Sharma14,Sharma15,Soo01}. The irony of this system is the self-protection of the quantum structures; no capping layer is needed to protect the quantum structures from external environment unlike the conventional quantum dot systems \cite{Brunner01}. A wetting layer of 1.1nm thick Ge was grown on top as shown in the high resolution cross-sectional transmission electron microscopy (XTEM) image in Fig. \ref{Fig1-core}(a) and Fig. \ref{Fig1-core}(b).

Photoemission measurements were carried out using a Gammadata Scienta R4000 WAL analyzer and monochromatic Al $K\alpha$ source at an energy resolution of 350 meV. All the measurements were carried out at a temperature, 20 K using an open cycle helium cryostat from Advance Research Systems. We exposed different parts of the IQH structures using controlled low energy Ar sputtering and annealing at 400 $^o$C. The exposed surface was identified by XTEM, EDX (energy dispersive analysis of $x$-rays) and SIMS (secondary ion mass spectrometry). In Fig. \ref{Fig1-core}(b), we show various surface terminations studied. For example, `s10(WL)' refers to 10 minutes sputtered surface revealing the wetting layer (WL). At s17, the base region of the IQHs is exposed (depth 1.1 nm). s42 and s72 are the mid regions of the IQHs having depth of 2.7 nm and 4.7 nm, respectively; these two regions expose the SiGe alloy within IQHs and the side interfaces between Si and IQHs. At s100, the tip of IQHs (depth of 6.6 nm) becomes visible.

Si 2$p$ and Ge 3$d$ core level spectra exhibit significantly different peak positions in the spectra from different locations as shown in Fig. \ref{Fig1-core}. The energy shift in Si 2$p$ spectra can be understood as follows. There are two types of Si - (i) Si(1) (no Ge neighbour) and (ii) Si(2) (Si in SiGe alloys of IQHs). The shift to lower binding energies can be attributed to the decrease in local potential due to hybridization with Ge \cite{Evjen32,dimen,prb-hike} and hence, indicates contribution from Si(2). The shift to higher binding energies at s17 and s100 indicates relatively larger contribution from Si(1) as expected after removal of wetting layer and the IQHs, respectively. The Ge 3$d$ peak, however, shifts by 200 meV towards higher binding energies from s10 to s17 whereas Si 2$p$ shifts by 50 meV. The shift in the same direction indicates Fermi level shift. Ge 3$d$ peak position remains almost unchanged at higher depth. The enhancement of Si 2$p$-Ge 3$d$ energy separation provide an evidence of hybridization between Si and Ge valence states and there is finite charge transfer between them due to variation of the relative concentration of Si/Ge \cite{Sharma14,Agarwal93}.

In addition, we discover new features at the lower binding energy side (21-28 eV) of Ge 3$d$ bulk peaks. The intensity of the features strongly depends on the location in the IQH structure and cannot be associated to impurities for the following reason. Bonding with carbon and/or oxygen (electronegative) leads to enhancement in binding energy as observed earlier; the Ge 3$d$ peaks in GeO$_{x}$ appear at higher binding energies \cite{Legoues89}. Our detailed characterization of the samples including elemental analysis does not show signature of impurity in the system. Photoemission spectroscopy being a highly surface sensitive technique, captures weak intensities due to adsorption of impurities on aging\cite{Bi2Se3-aging}. While weak intensity at 31.5 eV in Ge 3$d$ spectrum of s10 show signature of some surface oxygen, no such peak is present in other spectra having intense new features.

\begin{figure}
\includegraphics[scale=0.4]{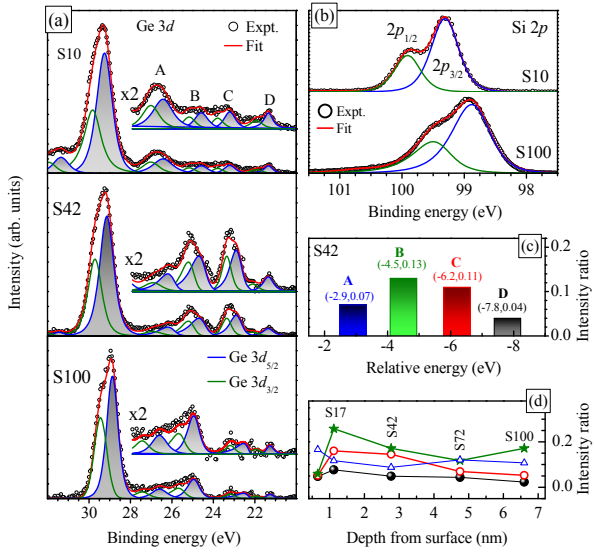}
\vspace{-12ex}
\caption{(a) Simulation of Ge 3$d$ spectra collected on s10, s42 and s100 surfaces. Solid line over the data points (symbols) represent the fit and other lines represent the component peaks. Inset shows the expanded view of the low binding energy region. (b) Simulation of Si 2$p$ spectra collected on S10 and S100 surfaces. Definition of lines are same as in (a). (c) Relative peak positions and intensities of the final state features in Ge 3$d$ spectrum of s42 surface with respect to the 3$d$ signals from Ge bulk. (d) Changes in relative intensities of different Ge 3$d$ low binding energy features with the variation of depth within IQH. Definition of colours is the same as in (c).}
\label{Fig2-fit}
\end{figure}

We simulated the experimental spectra to find the constituent features using least square error method following the selection rules and conservation rules associated to photoemission process. Si 2$p$ spectra could be simulated using two peaks representing spin-orbit split 2$p_{3/2}$ and 2$p_{1/2}$ signals with a spin-orbit splitting of 0.63$\pm$0.02 eV \cite{Schmeisser86}; a typical result is shown in Fig. \ref{Fig2-fit}(b). Fitting results of selected Ge 3$d$ core level spectra are shown in Fig. \ref{Fig2-fit}(a). Intense broad peak at 29 eV consists of Ge 3$d_{5/2}$ and 3$d_{3/2}$ signals (spin-orbit splitting of 0.58$\pm$0.02 eV) from bulk Ge \cite{Hollinger84}.

The features in 21-28 eV energy range are denoted by A, B, C and D. All the features could be captured well considering peak pairs possessing characteristics of Ge 3$d_{5/2}$ and 3$d_{3/2}$ signals. The position and intensities of A, B, C and D of s42 spectrum relative to the bulk Ge 3$d$ peak are shown in  Fig. \ref{Fig2-fit}(c). A, B, C and D are away from the main peak by about 2.8 eV, 4.6 eV, 6.5 eV and 7.8 eV, respectively. The intensities of the features at different depths from the sample surface are shown in Fig. \ref{Fig2-fit}(d). A and D are intense in s10 spectrum (the base region of IQHs). The intensity of A (blue line) exhibits a weak decrease and D (black line) exhibits a weak increase between s10 and s17. The intensity of D remains almost same at higher depth while the intensity of A exhibits weak non monotonic depth dependence. Thus, the peaks A and D are attributed to the final states having the origin related to the surface and interface of the wetting layer with the Si layers, respectively. The contribution of C is quite strong on S17 and S42 surfaces while the intensity of B is high in almost all the cases except S10.

\begin{figure}
\includegraphics[scale=0.4]{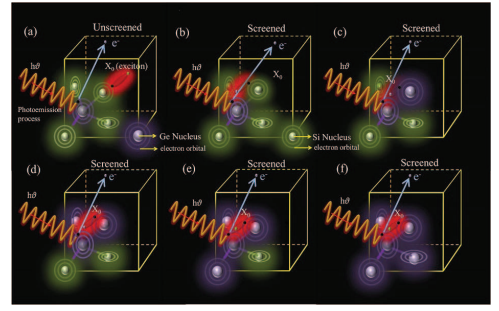}
\caption{Schematic of the core hole screening process by excitons. Few atoms are shown in each unit cell for clear visualisation of the process. Green and violate colours represent Si atoms and Ge atoms respectively and glow around the atoms represent corresponding electron clouds. Elongated red coloured glows represent excitons in the system.}
\label{Fig3-schem}
\end{figure}

The interaction of the photoexcited core hole and valence electrons, termed as core hole screening, leads to multiple features in core level spectrum; this is extensively studied using multiband Hubbard model \cite{RMP-Fujimori}. The well screened feature appear at lower binding energy relative to the poorly screened ones; the energy separation between the well screened and poorly screened features is often found to be of the order of 10 eV. Interestingly, it is observed that the extent of screening and the nature of the final state often gives rise to distinct features \cite{dimen,Ashish-PRL}. A good example of such a scenario is the final state features observed due to Zhange-Rice singlet appear about 1.5 eV lower in energy compared to the well screened feature \cite{Ashish-PRL}. The lowering of energy of the final state feature comes due to additional stabilization (energy scale $\sim$meV) as the ligand hole around the photoemission site created by the electron transfer for screening the core hole forms a bound state with another hole in the conduction band/neighbouring site \cite{Ashish-PRL,ZRS1,ZRS2,ZRS4}.

In semiconductors, core level spectra typically show poorly screened features as the valence electrons are localized – this is observed in Si and Ge core level spectra. However, the extended nature of the excitons can lead to significant screening of the core hole. The corresponding final state Hamiltonian can be expressed as,

$$H = \sum(\epsilon_c c^\dag c + \epsilon_p p^\dag p + \epsilon_e e^\dag e - \epsilon_{ex} p^\dag p e^\dag e) - \sum U_{ec}n_en_c$$
$$+ \sum(t_{ij}p_i^\dag p_j + t_{ij}^\prime e_i^\dag e_j + h.c.)$$

$p$, $e$ and $c$ represent hole in the valence band, electron in the conduction band and core hole due to photoemission, respectively. $\epsilon_{ex}$ is the binding energy of the exciton and $U_{ec}$ is the coulomb interaction strength between core hole and conduction electron. The filled valence band in this semiconductor is considered as an empty state, $|p^0>$. After photoemission, the possible states are $|p^0>$ and $|e^1L_p^1>$ – here the second state refers to an exciton, where the conduction electron resides at the photoemission site to screen the photoexcited core hole and it also has formed an excitonic state with the hole, $|L_p^1>$ in the neighbouring site. The energy difference between the two states will be a function of $U_{ec}$, $t$ and $\epsilon_{ex}$; the simplest form will be $\Delta E = \sqrt{[(\epsilon-U_{ec}-\epsilon_{ec})^2 + 4t^2]}$. Considering that the feature, D appears at 7.8 eV corresponds to the exciton having highest binding energy of 0.92 eV and the transfer energy is about 1 eV in these systems, the estimate of $U_{ec}$ will be about 5.3 eV, which is quite reasonable for such interactions. Clearly, the other features at relatively smaller energy separations can be captured using the parameters in this energy scale. While a detailed theoretical work is necessary to capture the complete physics behind this phenomena, the estimate provided here justifies the scenario of core hole screening by excitons with electronic interaction parameters consistent with the other experimental findings.

The qualitative scenario for various features is demonstrated in Fig. \ref{Fig3-schem}, where Ge 3$d$ core hole (black dot) forms at (0.25,0.25,0.25) and excitons ($X_0$: electron-hole bound pair) are shown by red glow. In Figs. \ref{Fig3-schem}(a), the exciton is not located at the photoemission site and hence the signal corresponds to the bulk feature. The exciton in \ref{Fig3-schem}(b) screens the core hole. It has been observed that excitons in Ge QDs have lower binding energy than the excitons in similarly shaped \& sized Si QDs \cite{Takagahara92} due to the lower bandgap. Therefore, the excitons associated with more numbers of Si atoms will be more strongly bound and the core hole screened state will be better stabilized. Thus, the corresponding photoemission feature will appear at the lowest binding energy, which is feature D. It is clear that the binding energies of various photoemission final states would be related to the number of Ge/Si atoms bonded to the Ge at the photoemission site as shown in Figs. \ref{Fig3-schem}(c-f). The mid region of IQH can be represented by the arrangements of Fig. \ref{Fig3-schem}(c) and Fig. \ref{Fig3-schem}(d), providing the features 'C' and 'B', respectively. Ge wet layer has atomic arrangement of the type shown in Fig. \ref{Fig3-schem}(e) having more Ge neighbours and can be associated with the feature 'A' in Ge 3$d$ spectra. The scenario of Fig. \ref{Fig3-schem}(f) is similar to bulk Ge and is less probable to find in our sample as supported by the observed compositional analysis of those IQHs \cite{Sharma14}. These descriptions beautifully match with the experimental observations.

In summary, we studied the core level spectra of Si/Ge quantum structures employing high resolution core level photoemission spectroscopy. The discovery of distinct features due to core hole screening by excitons provides a new direction of the core level spectroscopy; we demonstrate that the element specificity and surface sensitivity of CoLePES makes it a powerful direct probe for local excitons. The finding of local excitons in the inverted quantum huts embedded in Si-matrix are consistent with the type I photoluminescence \cite{Sharma15} and establish their candidature to engineer devices for optoelectronic applications and quantum communications.


M.K.S. and K.M. acknowledge financial assistance from the Department of Science and Technology, Govt. of India under J.C. Bose Fellowship program and the Department of Atomic Energy, Govt. of India. K. M. acknowledges financial support from Department of Atomic Energy, Govt. of India under DAE-SRC-OI award program.






\begin{thebibliography}{99}
\bibitem{growth}
The sample preparation procedure is as follows. Si (001) wafer was cleaned by removing the unwanted oxide layer, inorganic and organic contaminations following RCA cleaning method, which was followed by HF etching of top oxide layer on wafer. Immediately after the cleaning, the wafer was mounted in the load-lock chamber of molecular beam epitaxy (MBE) machine and shifted to preparation chamber where the base vacuum was maintained close to 1$\times$10$^{-11}$ Torr. In the preparation chamber, the wafer was kept at 900 $^{\circ}$C for 20 hours to remove further oxide layer grown on it during sample transfer through load-lock system. Later it had been transferred to ultra-high vacuum (UHV) growth chamber (base pressure around 10$^{-11}$ Torr) and heated up to 1100 $^{\circ}$C with a continuous flow of liquid N$_2$ to maintain the pressure, which opened up the dangling bonds of the top surface silicon atoms and provided the platform for epitaxial growth of Si and Ge layers on the surface. Subsequently, we have grown 100nm thick Si buffer layer at 750 $^{\circ}$C. This was followed by successive growth of 1.7nm Ge, 8nm Si and again 1.7nm Ge with deposition rate of 0.3 \AA/sec at 430 $^{\circ}$C. A wetting layer of 1.1nm thick Ge was deposited on top of the system.
\end{thebibliography}
\end{document}